\documentclass[a4paper]{article}

\usepackage{INTERSPEECH2022}
\usepackage{multicol}
\usepackage{multirow}
\usepackage{caption}
\usepackage{cleveref}
\usepackage{subcaption}
\usepackage{textcomp}
\newcommand{\algname}{\textsc{WavPrompt }}
\newcommand{\algnamens}{\textsc{WavPrompt}}

\title{\textsc{WavPrompt}: Towards Few-Shot Spoken Language Understanding with Frozen Language Models}
\name{Heting Gao$^1$, Junrui Ni$^1$, Kaizhi Qian$^2$, Yang Zhang$^2$, Shiyu Chang$^3$, Mark Hasegawa-Johnson$^1$}
\address{
  $^1$University of Illinois at Urbana-Champaign\\
  $^2$MIT-IBM Watson AI Lab
  $^3$University of California, Santa Barbara
}
\email{\{hgao17,junruin2,jhasegaw\}@illinois.edu, \{kqian, yang.zhang2\}@ibm.com, chang87@ucsb.edu}

\begin{document}

\maketitle
\begin{abstract}
    Large-scale auto-regressive language models pretrained on massive text have demonstrated their impressive ability to perform new natural language tasks with only a few text examples, without the need for fine-tuning. Recent studies further show that such a few-shot learning ability can be extended to the text-image setting by training an encoder to encode the images into embeddings functioning like the text embeddings of the language model. Interested in exploring the possibility of transferring the few-shot learning ability to the audio-text setting, we propose a novel speech understanding framework, \algnamens, where we finetune a wav2vec model to generate a sequence of audio embeddings understood by the language model.
    We show that \algname is a few-shot learner that can perform speech understanding tasks better than a na\"{i}ve text baseline. We conduct detailed ablation studies on different components and hyperparameters to empirically identify the best model configuration. In addition, we conduct a non-speech understanding experiment to show \algname can extract more information than just the transcriptions. Code is available at https://github.com/Hertin/WavPrompt
\end{abstract}

\section{Introduction}
\label{sec:intro}

Large-scale pretrained language models (PLM)~\cite{devlin2018bert,raffel2019exploring,radford2019language,brown2020language} have brought great success in natural language processing (NLP)~\cite{qiu2020pre}. Recently, researchers have discovered that PLMs demonstrate a strong capability for few-shot learning on many NLP tasks~\cite{brown2020language,zhao2021calibrate}. Specifically, if we feed to a language model a prefix containing several text-prompt-answer demonstrations of a task, as well as a new question, a language model can generate a decent answer to the new question upon seeing the prefix. Furthermore, by pretraining an image encoder to generate feature vectors that are meaningful to a PLM, the PLM can be given the ability to solve few-shot image understanding tasks~\cite{tsimpoukelli2021multimodal}.

We are therefore interested in whether such few-shot learning capabilities can generalize to spoken language understanding tasks as well. More specifically, our setting is as follows. Given a certain task, the task demonstrations are in the form of triplets containing 1) a speech utterance, 2) a text question/prompt, and 3) a text answer. We also have a new question that is in a similar form to the demonstrations but without an answer. Our goal is to convert the task demonstrations and the new question into a text prefix and feed it to a fixed language model, so that it can produce answers to the new question. Figure~\ref{fig:arch_all}(~\subref{fig:arch_eval}) shows an example, where the model is being taught to identify the gender of the person discussed in the speech utterance by seeing a few short demonstrations, each containing three components: first, a speech utterance (saying, e.g., `a woman in a red suit'), then a text prompt (`the speaker is describing a'), and finally the text answer (`woman'). Concatenated to the end of the training demonstrations is a question in a similar form but without the answer; the model is judged to perform correctly if it generates the correct answer (e.g., either `man' or `woman').

The main challenge of this task is to convert the speech into a form that can be accepted by the language model as the text prefix. One na\"{i}ve way is simply to convert the speech to text using an ASR, and then perform few-shot learning on the transcribed demos the same way as in NLP tasks. However, such a na\"{i}ve paradigm would propagate the errors in ASR to the language model, thereby undermining its few-shot learning performance. Also, this na\"{i}ve solution could not handle non-speech audio understanding tasks. We thus ask: are there better end-to-end solutions to speech understanding tasks?

In this paper, we propose \algnamens, an end-to-end few-shot learning framework for speech or audio understanding tasks. \algname consists of an audio encoder and a language model. The audio encoder is pretrained as part of an ASR, so that it learns to convert the speech in the demonstrations into embeddings digestible to the language model. After pretraining, the entire framework is frozen and ready to perform few-shot learning upon seeing the demonstrations.

We evaluate our model on speech classification tasks, and we can confirm that the zero-shot learning capabilities of fixed language models do generalize to simple speech understanding tasks. Furthermore, \algnamens, with its end-to-end pipeline, achieves a significant gain over the aforementioned na\"{i}ve ASR+NLP baseline. We further perform an extensive ablation study in search of the best hyperparameter settings. The findings of this paper can provide guidance and insights for research towards next-generation few-shot learning for speech and audio understanding.

% \section{Related Works}
% \subsection{Large Language Models as Few-Shot Learners}
% \subsection{Prefix Tuning and Prompt Tuning}
% \subsection{Multi-modal Prefix Tuning}

\section{Methods}
\subsection{Model Architecture}\label{sec:arch}

\begin{figure}[ht]
\setlength{\belowcaptionskip}{-2pt}
% \captionsetup{belowskip=0pt}
\begin{subfigure}{\linewidth}
  \centering
  \includegraphics[width=\linewidth, trim={0cm 1cm 0.55cm 1cm},clip]{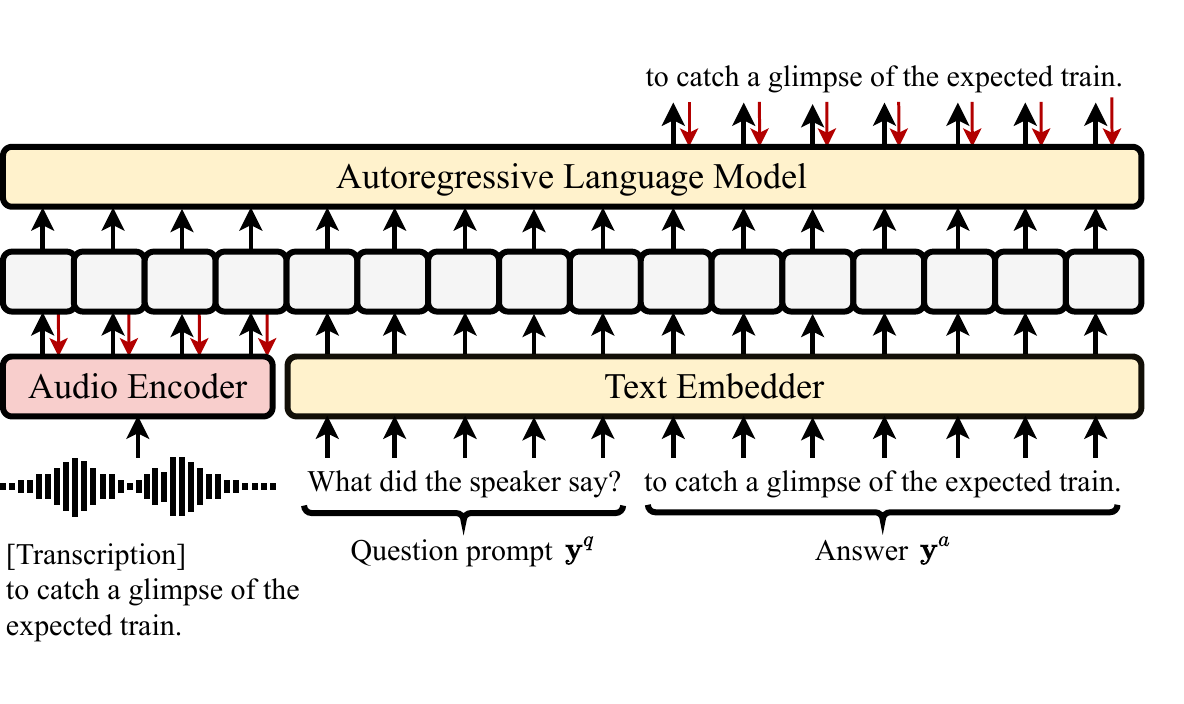}
  \caption{Interface of \algname during pretraining.}
  \label{fig:arch_pretrain}
\end{subfigure}

\begin{subfigure}{\linewidth}
  \centering
  \includegraphics[width=\linewidth, trim={0cm 1cm 0.55cm 0cm},clip]{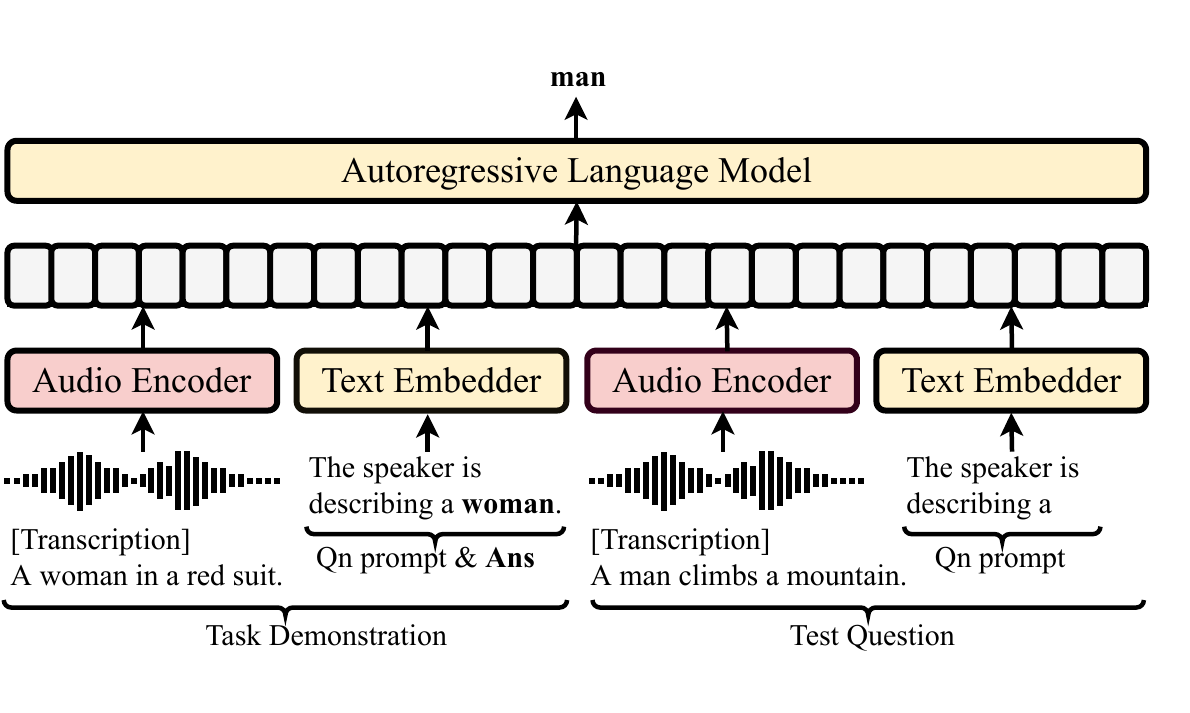}
  \caption{Interface of \algname during inference.}
  \label{fig:arch_eval}
\end{subfigure}
\caption{Interface of \algname}\label{fig:arch_all}
\end{figure}

Figure~\ref{fig:arch_all} shows the architecture of \algnamens, which consists of a pretrained audio encoder $f_\phi$, for which we use the wav2vec 2.0 base model~\cite{baevski2020wav2vec}, and a pretrained autoregressive language model, for which we use the GPT2~\cite{radford2019language}. The audio encoder $f_\phi$ encodes the speech audio $\mathbf x$ into continuous audio embeddings $\mathbf s=[s_1,s_2,...,s_m]=f_\phi(\mathbf x)$.  The language model contains a text embedder $h_\theta$ that converts the text $\mathbf y=[y_1,y_2,...,y_l]$ into a sequence of text embedings $\mathbf t=[t_1,t_2,...,t_n]=h_\theta(\mathbf y)$ and a transformer-based neural network $g_\theta$ that models the text distribution $p(\mathbf y)$ as 
\begin{align*}
    \log p(\mathbf y)=\sum_{i=1}^n \log p(t_i|t_1,...,t_{i-1})=\sum_{i=1}^n g_\theta(t_1,...,t_{i-1})_{t_i}
\end{align*}
% The language model contains a text embedder $h_\theta$ that converts the text $y$ into a sequence of text embedings $t=[t_1,t_2,...,t_n]$. 

\subsection{Downsampling Layer}\label{sec:arch_ds}
The wav2vec audio encoder takes $16$ kHz audios and extracts feature vectors at a frequency of $50$ Hz. A simple calculation gives us that the LibriSpeech ASR corpus~\cite{panayotov2015librispeech} has an average of $2.7$ words per second and $4.9$ tokens per second using GPT2's tokenizer. This means the text embedding vectors have a frequency of roughly $5$ Hz, which is only $10\%$ the rate of the audio embeddings. Therefore we append a downsampling layer after the audio encoder to reduce the rate of audio embeddings so that the rate of the audio embedding can better match that of the text embeddings. %We experiment with different downsampling rates to empirically study how the downsampling rate affects the performance. 

\subsection{Speech Recognition Pretraining}
We pretrain \algname as an ASR, using the 100-hour train-clean split of the LibriSpeech ASR corpus~\cite{panayotov2015librispeech}. We also create 5-hour and 10-hour pretraining datasets by sampling from the 100-hour split to simulate low resource conditions. 

% Similar to the multimodel prefix paper~\cite{Li2021PrefixTuningOC}, 
We keep the language model fixed and only update the audio encoder during pretraining. An overview of the pretraining interface is shown in Figure~\ref{fig:arch_all}(\subref{fig:arch_pretrain}), where the red arrows denote the back-propagation of the gradients.
The audio embeddings $\mathbf s$, together with the text embeddings $\mathbf t^q=[t^q_1,t^q_2,...,t^q_n]$ of the question prompt $\mathbf y^q$ are fed to the language model so that the language model models the probability of the answer $\mathbf y^a$ conditioned on the audio and the question prompt as
\begin{align*}
    \log p(\mathbf y^a|\mathbf x,\mathbf y^q)&=\sum_{i=1}^l p(t^a_i|\mathbf s,\mathbf t^q,t^a_1,...,,t^a_{i-1})\\
    &=\sum_{i=1}^l g_\theta(s_1,...,s_m,t^q_1,...,t^q_n,t^a_1,...,t^a_{i-1})_{t^a_i}
\end{align*}
We use the question `what did the speaker say?' as a prompt during pretraining.
% We use the text question, `what did the speaker say?' together with the audio embedding to prompt the language model to generate correct transcriptions to the audios. The question prompt is converted to a sequence of text embeddings using the GPT2's embedding code book. The audio embeddings from the encoder concatenated with the prompt embeddings are input to the GPT2 to condition the transcription generation. During pretraining, the GPT2 language model is kept frozen and only the wav2vec audio encoder is updated.

\subsection{Few-Shot Evaluation}
We evaluate \algname on few-shot binary classification tasks. During evaluation, we do not update the model parameters.
Instead, the model is given a single prompt sequence that contains from 0 to 10 demonstrations of a new task, followed by a question that it must answer using the form specified in the demonstrations.
%The answers is the text labels that are semantically related to the speech content and \algname need to predict the correct label given the testing samples and the few-shot examples. 
An illustration of the inference interface during evaluation is shown in Figure~~\ref{fig:arch_all}(\subref{fig:arch_eval}). 
As shown, the question is usually a sentence with a gap at the end; \algname must fill the  gap based on the content of the audio.
Unlike the setting in~\cite{Li2021PrefixTuningOC}, we restrict each task to a finite output space (either two or nine possible answers), so that the accuracy of the few-shot learner can be meaningfully compared to chance performance.
%Note that this setting is simpler than that in the multimodal prefix tuning paper~\cite{Li2021PrefixTuningOC} as we constrain 
%each of our few-shot speech understanding 
%the output space to only two labels. We choose this setting because in preliminary experiments we find it hard to have the GPT2 model to generate the correct label, possibly because the language model we use has only $1.4\%$ of parameters of the one in~\cite{Li2021PrefixTuningOC} and is therefore less powerful. However, the evaluation results still tell us how to prompt the frozen language model with audio embeddings to answer multimodal question in the few-shot setting. 

% \begin{figure}[ht]
%   \centering
%   \includegraphics[width=\linewidth, trim={0cm 1cm 0cm 1cm},clip]{LaTeX/IEEEtran/fig/wavprompt.drawio.pdf}
%   \caption{Interface of \algname during inference.}
%   \label{fig:arch_eval}
% \end{figure}

\subsection{Calibration}
During the evaluation, we find that the performances of our models are not stable. We therefore implement the calibration technique reported by~\cite{zhao2021calibrate} to reduce the bias introduced by the language models. We empirically find the calibration brings improvement to the classification accuracy in most cases.

\section{Experiments}
\begin{figure*}[t]
  \centering
  \includegraphics[width=\linewidth, trim={0cm 0.5cm 0cm 0cm},clip]{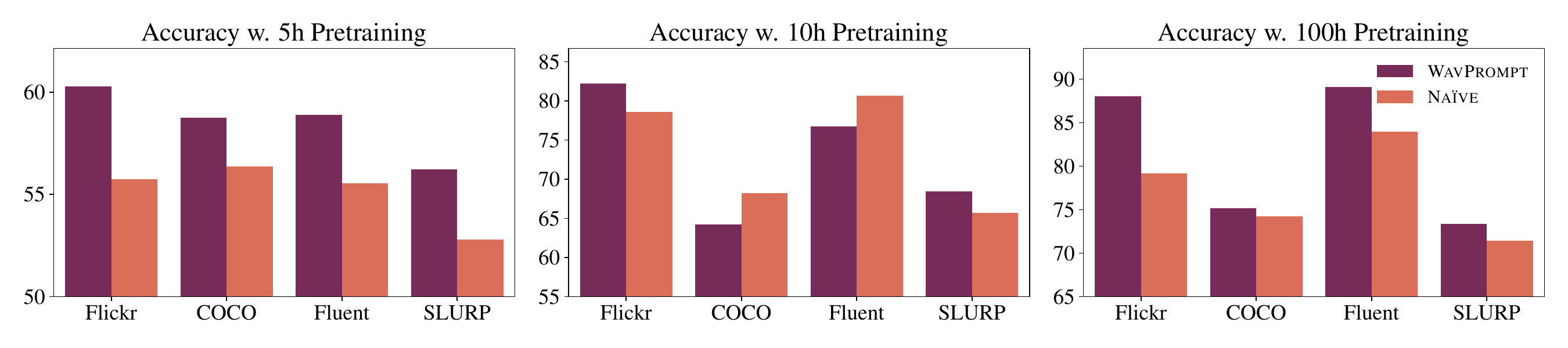}
  \caption{Results of speech understanding tasks.}
  \label{fig:ne2e}
\end{figure*}
\subsection{Datasets}
We evaluate \algname on four speech datasets: Flickr8k Audio Caption Corpus (Flickr8k)~\cite{Harwath2015DeepMS}, Fluent Speech Commands corpus (Fluent)~\cite{lugosch19_interspeech}, Spoken Language Understanding Resource Package (SLURP)~\cite{bastianelli-etal-2020-slurp} and SpokenCOCO Audio Caption corpus (SpokenCOCO)~\cite{Hsu2021TextFreeIS}. In addtion, we evaluate \algname on a non-speech dataset: Environmental Sound Classification (ESC50)~\cite{piczak2015dataset}. Brief introductions and the preprocessing steps of the dataset are as follows.

\vspace{2mm}
\noindent\textbf{Flickr8k}\quad
The Flickr8k Audio Caption Corpus is an extension to the Flickr8k Dataset~\cite{hodosh2013framing} containing 40,000 spoken captions of 8,000 natural images. We drop the images and only use the spoken caption audios and their transcripts. We then randomly sample 2000 captions and manually assign four sets of labels to the captions. We form man-woman label set by assigning `man' and `woman' labels to the captions that contain either only `man' or only `woman' words respectively. We create the male-female label set by replacing `man' and `woman' labels with `male' and `female' labels. We repeat the procedure to form black-white and dark-light label sets, but we additionally drop those that are not describing the color of the clothes. The resulting subset contains around 400 man-woman and male-female labeled samples and around 70 black-white and dark-light labeled samples. We use `the speaker is describing a person in' as the question prompt for the color labels and `The speaker is describing a' for the gender labels. This dataset is mainly used to probe if the model can capture semantic relations between word-pairs. 

\vspace{2mm}
\noindent\textbf{SpokenCOCO}\quad
Similar to the Flickr dataset, the SpokenCOCO Audio Caption dataset contains approximately 600,000 spoken captions describing the images in the Microsoft COCO (MSCOCO) dataset~\cite{lin2014microsoft}. The MSCOCO dataset classifies the images using 12 super-category labels, which we use as the labels of the spoken captions. During evalution, we ask the model to discern between the `vehicle' labels and the rest of labels, forming a total of 11 classification tasks. We use `The speaker is describing' as the question prompt.

\vspace{2mm}
\noindent\textbf{Fluent}\quad
The Fluent dataset contains spoken commands that interact with smart devices, such as `play the song' and `increase the volume.' Each command is labeled with action, object and location. We 
%combine action and object labels to form topic label the command is related to.
define topic labels to be the
%The topic label is 
same as the object label most of the time, except that when the action is `change language,' the topic is set to `language' instead of the actual language name. We use `The topic is' as the question prompt.

\vspace{2mm}
\noindent\textbf{SLURP}\quad
Similar to Fluent dataset, SLURP is also a dataset for spoken language understanding that contains human interaction with home assistants from 18 different domains. We select five domains: `music', `weather', `news',  `email' and `play' and form ten domain pairs for our model to perform binary classification. We use `This is a scenario of' as the question prompt.

\vspace{2mm}
\noindent\textbf{ESC50}\quad
The ESC50 dataset contains 2000 environmental audio recordings including animal sounds, human non-speech sounds, natural soundscapes, domestic and urban noises, etc. We use the sound label as groundtruth text and pretrain additional \algname models on the 100-hour Librispeech dataset and ESC50 dataset for ASR and environment sound classification tasks simultaneously. During pretraining, we prompt the model with `What did the speaker say?' for the ASR task and `What sound is this?' for the environment sound classification task. We test the model on a subset of the training set that only contains sounds of nine animals: dog, cat, bird, sheep, cow, pig, rooster, hen and frog. During testing we assign a distinct verb to each of the nine animals: barks, chirps, bleats, meows, moos, snorts, crows, clucks and croaks. \algname needs to predict the correct verb given the animal sound and a few examples. We use `=\texttt{>}' as the question prompt during evaluation.

\subsection{Experiment Setup and Baseline}
We modify the fairseq~\cite{ott2019fairseq} training pipeline for our experiment. We use the wav2vec 2.0 base model implemented in fairseq as audio encoder and the GPT2 model with 117 million parameters implemented in Huggingface~\cite{wolf-etal-2020-transformers} as our language model.

For the speech classification tasks, we pretrain a total of 15 \algname models with five downsampling rates (2, 4, 8, 16, 32) under three resource conditions (5, 10 and 100 hours of Librispeech data). For the non-speech classification tasks, we pretrain five \algname models with five downsampling rates (2, 4, 8, 16, 32) using 100 hours of Librispeech data. 

% We evaluate two architectures: the proposed architecture, which uses \algname models directly, and the na\"{i}ve baseline where prompted with the audio embedding and the pretraining prompt, \algname transcribes the audio into text which is then fed to the language model for classification. 

During evaluation, we randomly sample several samples along with their correct labels from the test set as shots. The shots are converted to embeddings and are prepended to the question prompt embeddings. We sample 250 samples from the rest of the test set to form an evaluation batch and drop samples from the class containing more samples to evenly balance the class labels in the batch. As a result, a binary classification accuracy greater than 50\% is better than chance. We sample five batches with different random seeds. The classification accuracy we report is the average accuracy over the five batches.

We compare the \algname with the \textsc{Na\"{i}ve} baseline mentioned in Section~\ref{sec:intro}, which converts the speech into text and performs few-shot learning using the transcribed text. Specifically, \textsc{Na\"{i}ve} uses the same model as \algnamens. It performs few-shot learning via two steps. First, the speech is converted into text using an ASR. To achieve this, we use the pretrained \algname as an ASR by prompting the language model with the audio embedding and the pretraining question `what did the speaker say?'. Second, to perform few-shot learning, we prompt the language model with the transcribed text embeddings instead of audio embeddings. In other words, the only difference between \algname and \textsc{Na\"{i}ve} is that the audio embeddings are used in the prompt in the former, whereas the transcribed text embeddings are used in the latter.

\subsection{Results on the Speech Understanding Tasks}\label{sec:ne2e}
% \subsection{End-to-End versus Non-End-to-End}

% We compare the performance of the end-to-end (E2E) architecture (our model) and the non-end-to-end (NE2E) architecture, where prompted using the audio embedding and pretraining prompt, the model transcribes the audio into text and then the text are input to the language model.
Figure~\ref{fig:ne2e} shows the results on the four speech understanding tasks. To factor out the influence of numbers of shots, we use the best accuracy achieved over all numbers of shots to represent the model's performance on individual pairs of labels,
for both \algname and \textsc{Na\"{i}ve}. 
We average the accuracy over all label pairs in a dataset as the overall accuracy. 
We select the best-calibrated model among all the downsampling rates for both \algname and the \textsc{Na\"{i}ve} to make a fair comparison. We compute the overall accuracy of the model across four speech understanding datasets under three resource conditions. 

As can be observed, both algorithms can achieve an accuracy significantly above chance, which confirms that language models can perform zero-shot learning on speech understanding tasks. Also, the performance increases as the pretraining dataset size increases. Finally, \algname consistently outperforms \textsc{Na\"{i}ve} in almost all cases across datasets and across resource conditions, which verifies the advantage of training an end-to-end framework.

% \begin{table}[th]
%   \caption{Classification accuracy between end-to-end and non-end-to-end architecture.}
%   \label{tab:e2e}
%   \centering
%   \begin{tabular}{ l l l| l l| l l}
%     \toprule
% 	    &\multicolumn{2}{c}{5h}	&\multicolumn{2}{c}{10h} &\multicolumn{2}{c}{100h}	\\	
% 	    & E2E & NE2E & E2E & NE2E & E2E & NE2E \\\midrule	
% % Flickr	&\bf57.99	&52.64	&\bf82.21	&78.62	&\bf85.73	&79.14 \\
% % Fluent	&\bf59.22	&54.80	&76.77	&\bf80.66	&\bf90.04	&83.94 \\
% % SLURP	&\bf56.23	&52.69	&\bf68.47	&65.68	&\bf73.37	&71.01 \\
% Flickr	&\bf60.27	&55.74	&\bf82.21	&78.62	&\bf88.03	&79.14 \\
% COCO	&\bf58.74	&56.37	&64.24	&\bf68.21	&\bf75.15	&74.24 \\
% Fluent	&\bf58.88	&55.54	&\bf79.80	&79.76	&\bf89.11	&83.94 \\
% SLURP	&\bf56.23	&52.78	&\bf68.47	&65.68	&\bf73.37	&71.42 \\
%     \bottomrule
%   \end{tabular}
% \end{table}

\subsection{Ablation Studies}\label{sec:ablation}
% We conduct ablation studies on the effects of the downsampling rate, the calibration and the number of shots.
% to see how they affect the classification accuracy.
% \subsection{how do the speech only system compare with text system under different amount of resource}
% \subsection{explore what setting do the system works best}

% \subsection{generalization to other speech dataset}
% \subsection{rule to select best hyperparameter setting}

% \subsubsection{Downsampling Rate}

% The same model prompted with different numbers of shots has different classification accuracy. 

\noindent\textbf{Downsampling Rate}\quad\label{sec:expdsrate}
We use the best accuracy over all numbers of shots to represent the model performance as in Sec~\ref{sec:ne2e}. We average the best accuracy over all pairs of labels in each dataset and present the results in Table~\ref{tab:best_dsrate}. The results are consistent across datasets, suggesting that a downsampling rate of $8$ gives the best accuracy when the model is pretrained using $10$ or more hours of data and a downsampling rate of $4$ gives better accuracy when the model is trained using $5$ hours of data. The best downsampling rate being $8$ is expected as it produces the audio embeddings at a rate closest to that of the text embeddings as discussed in Sec~\ref{sec:arch_ds}.
\begin{table}[t]
  \caption{Classification accuracy across downsampling rates.}
  \label{tab:best_dsrate}
  \centering
  \footnotesize{
  \begin{tabular}{ l@{\hspace{1.5\tabcolsep}} c@{\hspace{1.5\tabcolsep}} c@{\hspace{1.5\tabcolsep}} c@{\hspace{1.5\tabcolsep}} c@{\hspace{1.5\tabcolsep}} c@{\hspace{1.5\tabcolsep}} c}
    \toprule

	 &Dataset&2	    &4	    &8	    &16	    &32  \\\midrule\midrule
5h & Flickr & 57.82 & 57.99 & \bf60.27 & 56.63 & 52.28 \\
 & COCO & 55.08 & \bf58.74 & 56.92 & 56.62 & 53.79 \\
 & Fluent & 55.54 & \bf58.88 & 57.4 & 57.35 & 53.32 \\
 & SLURP & 54.40 & \bf56.23 & 53.94 & 54.85 & 51.97 \\\midrule
10h & Flickr & 64.91 & 65.77 & \bf82.21 & 79.23 & 56.84 \\
 & COCO & 59.56 & 59.18 & \bf64.24 & 54.99 & 55.84 \\
 & Fluent & 64.21 & 72.01 & \bf79.80 & 64.58 & 53.41 \\
 & SLURP & 55.95 & 57.00 & \bf68.47 & 60.53 & 54.16 \\\midrule
100h & Flickr & 82.61 & \bf88.03 & 85.73 & 79.96 & 79.40 \\
 & COCO & 68.52 & 67.68 & \bf75.15 & 67.77 & 65.72 \\
 & Fluent & 82.47 & 87.36 & \bf89.11 & 81.12 & 83.90 \\
 & SLURP & 72.05 & 72.07 & \bf73.37 & 68.69 & 68.63 \\
%  \midrule\midrule
% NE2E &Dataset&2	    &4	    &8	    &16	    &32  \\\midrule
% 5h & Flickr & 54.81 & 52.64 & 55.5 & \bf55.74 & 54.43 \\
%  & COCO & 55.3 & 55.96 & \bf56.37 & 54.9 & 53.77 \\
%  & Fluent & 55.01 & \bf55.54 & 55.13 & 53.38 & 53.66 \\
%  & SLURP & 52.31 & 52.69 & 51.28 & \bf52.78 & 52.63 \\\midrule
% 10h & Flickr & 57.25 & 61.95 & \bf78.62 & 77.78 & 55.77 \\
%  & COCO & 61.33 & 61.0 & \bf68.21 & 61.09 & 54.53 \\
%  & Fluent & 57.98 & 60.97 & \bf79.76 & 70.04 & 57.13 \\
%  & SLURP & 53.35 & 53.35 & \bf65.68 & 59.93 & 53.06 \\\midrule
% 100h & Flickr & 77.97 & 78.33 & \bf79.14 & 77.93 & 78.51 \\
%  & COCO & 72.35 & 73.06 & \bf74.24 & 74.12 & 71.62 \\
%  & Fluent & 81.08 & 83.72 & \bf83.94 & 82.04 & 80.49 \\
%  & SLURP & 71.25 & \bf71.42 & 71.01 & 70.2 & 69.29 \\
    \bottomrule
  \end{tabular}
  }
\end{table}
		
% \subsubsection{Calibration}
\vspace{2mm}
\noindent\textbf{Calibration}\quad
We compare the classification accuracy with calibration versus without calibration using the best downsampling rate obtained in Table~\ref{tab:best_dsrate}. For each dataset we average the best classification accuracy over all label pairs for both the model with calibration and without calibration. The results are presented in Table~\ref{tab:calibration}. Almost in every case the model with calibration outperforms that without calibration by a large margin, suggesting necessity of the calibrating the PLM.
The only exception occurs in the model pretrained using $10$ hours of data and tested on Flickr dataset, but even in that case the accuracies are comparable.
\begin{table}[th]
  \caption{Classification accuracy between the model with calibration and without calibration denoted by `Cali' and `NCali' respectively.}
  \label{tab:calibration}
  \centering
  \footnotesize{
  \begin{tabular}{l c c c c c c}
%   { l@{\hspace{1.5\tabcolsep}} c@{\hspace{1.5\tabcolsep}} c@{\hspace{1.5\tabcolsep}} c@{\hspace{1.5\tabcolsep}} c@{\hspace{1.5\tabcolsep}} c@{\hspace{1.5\tabcolsep}} c}
    \toprule
	\multirow{2}{*}{} &\multicolumn{2}{c}{5h}	&\multicolumn{2}{c}{10h} &\multicolumn{2}{c}{100h}	\\	
	    \cmidrule(lr){2-3}\cmidrule(lr){4-5}\cmidrule(lr){6-7}
	    & Cali & NCali & Cali & NCali & Cali & NCali \\\midrule	\midrule
% Flickr	&\bf57.99	&54.81	&82.21	&\bf82.42	&\bf85.73	&82.69 \\
% Fluent	&\bf59.22	&58.50	&\bf76.77	&65.60	&\bf90.04	&87.54 \\
% SLURP	&\bf56.23	&55.09	&\bf68.47	&66.04	&\bf73.37	&69.52 \\
Flickr	&\bf60.27	&54.89	&82.21	&\bf82.42	&\bf88.03	&84.78 \\
COCO	&\bf58.74	&51.72	&\bf64.24	&59.15	&\bf75.15	&68.95 \\
Fluent	&\bf58.88	&58.50	&\bf79.80	&66.85	&\bf89.11	&87.54 \\
SLURP	&\bf56.23	&55.09	&\bf68.47	&66.04	&\bf73.37	&69.52 \\
    \bottomrule
  \end{tabular}
  }
\end{table}

\begin{figure}[t]
  \centering
  \includegraphics[width=\linewidth, trim={0cm 0cm 0.3cm 0.3cm},clip]{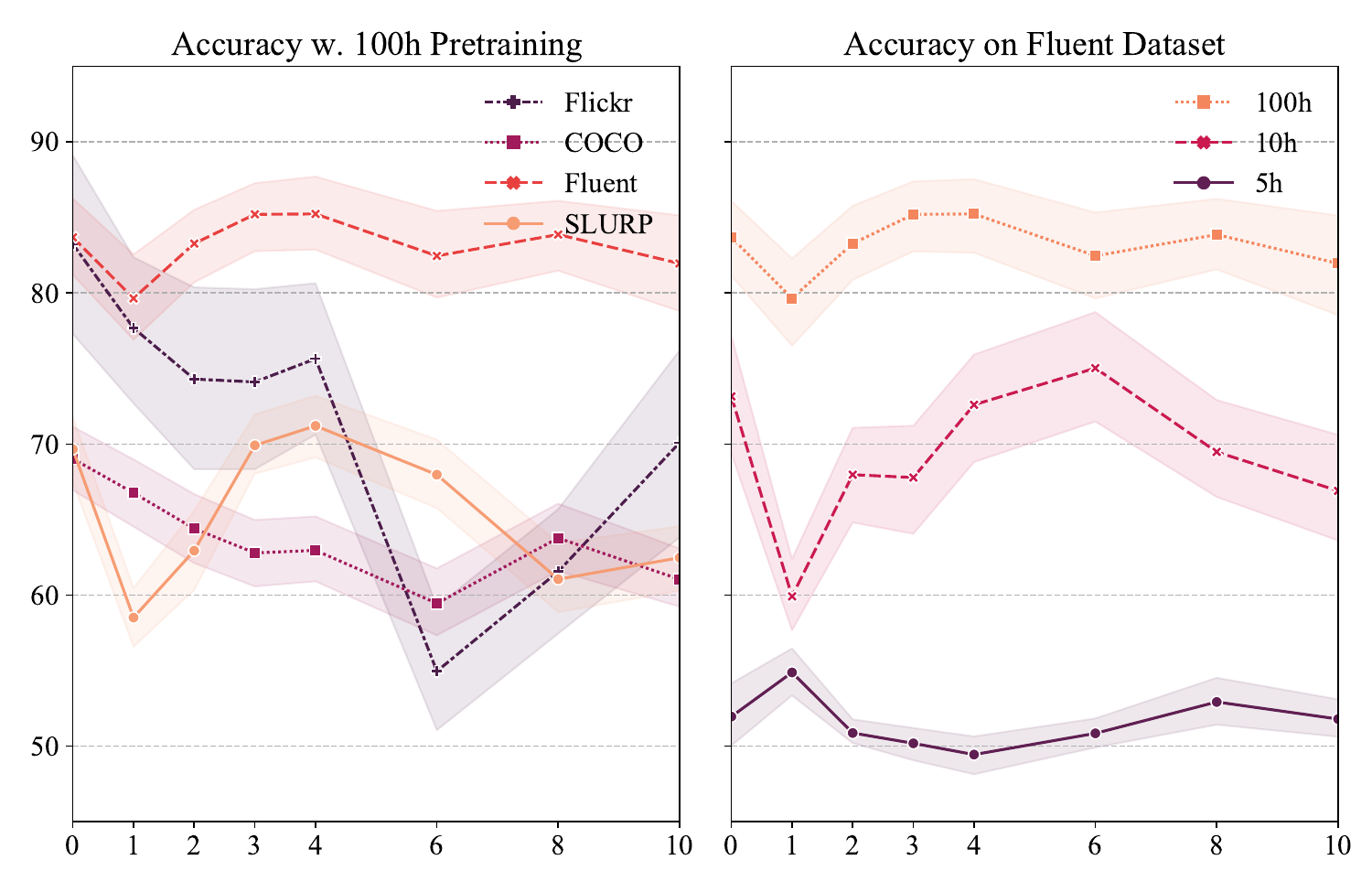}
  \caption{Classification accuracy versus number of shots.  Shaded region is $\pm 1$ standard deviations.}
  \label{fig:n_shot}
\end{figure}

% \subsubsection{Number of Shots}
\vspace{2mm}
\noindent\textbf{Number of Shots}\quad
To study the effect of the number of shots, we plot the classification accuracy across different datasets under 100-hour LibriSpeech dataset in the left subplot of Figure~\ref{fig:n_shot} and plot the accuracy across different resource conditions on Fluent dataset in the right subplot of Figure~\ref{fig:n_shot}. Although the accuracy curves exhibit different patterns across different datasets and different resource conditions, we observe that there usually exists two 
%peaks, the first occurring at the zero shot and the second occurring at four or six shots. 
peaks: one with zero demonstration examples, one with four to six demonstrations.
In Flickr and COCO experiments, zero-shot gives the best performance and increasing number of shots does not bring any benefits. One possible explanation is that the
Flickr and COCO datasets are simpler than the Fluent and SLURP datasets, in the sense that the class labels or their near synonyms occur directly in the speech; since the model has been pretrained as an ASR, the neurosymbolic representations of these answers may be already activated in the language model, so that the extra activation provided by the question is sufficient to generate a correct answer, even with zero demonstration examples.
%and providing examples does not give more information in this case. 
In Fluent and SLURP experiments, increasing shots to four or six yields the best accuracy but further increasing shots downgrades the performance. Using a larger language model might result in a more consistent pattern, which we leave as future work to explore.

\subsection{Generalizing to Non-Speech Tasks}
We additionally conduct a classification experiment using ESC50, a non-speech dataset.
Prompted with a few examples, \algname needs to predict the correct verb corresponding to the animal that makes the non-speech sound. We also provide a text baseline that replaces audio embedding with the text embedding of the animal's name.  As in previous sessions, we use the best accuracy across number of shots to represent the model's performance, for both \algname and the baseline.  The results are presented in the Table~\ref{tab:animal}. 

We observe that the classification accuracies are all better than chance, which is $11.11\%$ for a nine-way classification, and the best performing \algname with a downsampling rate of 8 is slightly better than the text baseline. These results show that \algname is able to extract information from non-speech audio and then leverage commonsense knowledge from its pretrained language model to solve problems.
%, suggesting its potential to be a versatile model more than a speech transcriber.
\begin{table}[th]
  \caption{Classification accuracy across downsampling rates on ESC50 dataset.}
  \label{tab:animal}
  \centering
  \footnotesize{
  \begin{tabular}{l c c c c c c }
%   {l@{\hspace{1.5\tabcolsep}} c@{\hspace{1.5\tabcolsep}} c@{\hspace{1.5\tabcolsep}} c@{\hspace{1.5\tabcolsep}} c@{\hspace{1.5\tabcolsep}} c@{\hspace{1.5\tabcolsep}} c }
    \toprule
     &2	    &4      &8	        &16	    &32     & text \\\midrule\midrule
ESC50&38.11	& 31.04 &\bf43.50	&32.89	&24.26  & 42.22\\
    \bottomrule
  \end{tabular}
  }
\end{table}

\section{Conclusions}
In this paper, we propose a novel speech understanding framework, \algnamens, and show that \algname is a few-shot learner that can perform both speech and non-speech understanding tasks better than a na\"{i}ve ASR baseline. We conduct detailed ablation studies on different components and hyperparameters to empirically identify the best model configuration. 
% In addtion, we conduct a non-speech understanding experiment to show \algname can extract information more than just the transcriptions.
\section{Acknowledgements}
This work is supported by IBM-UIUC Center for Cognitive Computing Systems Research (C3SR).

\bibliographystyle{IEEEtran}

\bibliography{mybib}

\end{document}